\documentclass{epl}

\title{Measurement of entropy production rate in compressible turbulence.}
\shorttitle{Entropy rate in compressible turbulence.}
\author{M. M. Bandi \inst{1,2} \and W. I. Goldburg \inst{1} \and J. R. Cressman Jr. \inst{3}}

\institute{
  \inst{1} Dept. of Physics \& Astronomy, University of Pittsburgh, Pittsburgh, PA 15260, USA.\\
  \inst{2} Center for Nonlinear Studies and Condensed Matter \& Thermal Physics Division, Los Alamos National Laboratory, Los Alamos, NM 87545, USA.\\
  \inst{3} Krasnow Institute, George Mason University, Fairfax, VA 22030, USA.
}
\pacs{47.27.-i}{Turbulent Flows}
\pacs{47.27.ed}{Dynamical Systems Approaches}
\pacs{47.52.+j}{Chaos in fluid dynamics}

\begin{document}

\maketitle

\begin{abstract}
The rate of change of entropy $\dot S$ is measured  for a system of particles floating on the surface of a fluid maintained in a turbulent steady state.  The resulting coagulation of the floaters allows one to relate $\dot S$ to the the velocity divergence and to the Lyapunov exponents characterizing the behavior of this system. The quantities measured from experiments and simulations are found to agree well with the theoretical predictions.
\end{abstract}

\section{Introduction}

It is common to probe the velocity field of a turbulent fluid by measuring velocity differences or correlation functions across various spatial or temporal scales. Much is also learned from the probability density functions themselves. Here the dynamics of the flow is investigated by measuring the system's rate of entropy change. In a sense that will be elaborated upon, the entropy rate measured here is zero unless the flow is compressible. It is not required, however, that the flow speeds be very large. Indeed the present measurements are made at only moderate Reynolds number \cite{boffetta2004}.

The entropy rate measurements are made by recording the inhomogeneous distribution of particles that float on the surface of a strongly turbulent tank of water \cite{robNJP2004}. The velocity of these floaters is completely controlled by the turbulent velocity field beneath them. One fully expects that this three-dimensional (3D) flow is governed by the (incompressible) Navier-Stokes equation (NSE).

The entropy $S(t)$ is defined in terms of the floater concentration at the fluid's surface. From measurements of the velocity and floater concentration fields, one extracts the entropy rate $dS/dt \equiv \dot S$, at the surface. By photographically tracking the motion of floaters, $\dot S$ is measured under transient conditions and after the system has reached a steady state. In interpreting the measurements, it is assumed that the flow underlying the dynamics of the Lagrangian tracers, which are assumed to follow the streamlines (ie. the inertial effects can be neglected), is maximally chaotic and therefore admits SRB (Sinai-Ruelle-Bowen) statistics \cite{boffetta2004}. If so, $\dot S$ is equal to the sum of the Lyapunov exponents characterizing the underlying flow.

Fortunately there exists a direct numerical simulation of the NSE that yields these exponents and hence their sum. Thus the predictions, based on chaos theory, can be directly confronted with simulations having as a starting point, the NSE itself.  To our knowledge these are the first measurements of the rate of entropy change for a many body system driven far from thermal equilibrium. Other recent studies \cite{goldenfeld2006,grisha2006} give further evidence that our understanding of non-equilibrium dynamical processes maybe illuminated by invoking ideas usually associated with equilibrium statistical mechanics and critical phenomena.

The floaters sample the velocity field at the surface of water, which is incompressible.  For them an initially uniformly distributed array of particles will subsequently tend to cluster under the influence of the underlying turbulent fluid motion, as shown in fig.~\ref{fourclouds}, thus reducing the total entropy of the floaters, defined here as:

\begin{equation}
\label{S}
S(t) = -\int~d{\bf r}~n({\bf r},t)\ln~n({\bf r},t).
\end{equation}
Here $n({\bf r},t)$ is the local concentration of particles on the surface. The integral is over an area spanned by an overhead camera that records the motion of floaters. The dimensions of the water tank  are large compared to the largest scales of turbulent flow.  To understand the statistical mechanics of floaters, one need not be concerned with details of the turbulent flow that drives them ~\cite{robNJP2004,bandi2006}.

 Falkovich and Fouxon ~\cite{grishaNJP2004} (FF) have shown that the entropy production rate ${\dot S}$ of compressible systems, like the floating particles in this experiment,  is nonzero.  The floaters  ``live" on the surface and thus have a two-dimensional velocity divergence $ \omega({\bf r},t) \equiv {\vec\nabla_2}  \cdot {\vec v}({\bf r},t)$.  The fluid, being incompressible everywhere, including the surface, tracer particles (specific gravity 0.25) obey the equation $\vec\nabla_{2} \cdot \vec v = \partial_x v_{x} + \partial_y v_{y} = - \partial_z v_{z} \neq 0$. The finite compressibility of the floating particle system is the origin of the nonzero value of ${\dot S}$.

The entropy rate can be written as the sum of two terms ~\cite{grishaNJP2004}:
\begin{equation}
\label{sdot1}
{\dot S} = \int_{A}~d{\bf r}~n({\bf r},t)\omega({\bf r},t) + \int_{B}~n({\bf r},t) \ln~n({\bf r},t){\bf v} \cdot d{\bf S}
\end{equation}
The first term is an integral over the area (area term) spanned by a camera which records the motion of floaters; the second term is a line integral around the periphery (boundary term). It takes into account particles that escape from the camera's field of view. Eq.~\ref{sdot1} is derived from eq.~\ref{S} by applying the condition for mass conservation which translates to conservation of particles in this experiment ($\partial_t n({\bf r},t) + \nabla_2 \cdot (n({\bf r},t){\bf v}({\bf r},t)) = 0$). FF assume that the boundary term is zero and work exclusively with the area term. Under the assumption that the floaters admit SRB statistics, they show that the first term on the right is the integral of the temporal correlation function of the lagrangian velocity divergence with a negative prefactor, as well as the sum of the system's Lyapunov exponents.

\begin{equation}
\label{sdot2}
{\dot S}_A = -\int~d\tau~\langle \omega({\bf r},t)\omega({\bf r},t+\tau) \rangle = \lambda_1 + \lambda_2
\end{equation}

Here {\bf r} is the starting point of the trajectories, and $\langle...\rangle$ represent an average over both $t$ and ${\bf r}$. This correlation function holds only if the points $r$ are uniformly distributed in space at the initial time $t$. $\lambda_1$ and $\lambda_2$ are the two lyapunov exponents characterizing the system's chaotic evolution with the convention that $\lambda_1 > \lambda_2$. Reported here are measurements of both the area and boundary term contributions to ${\dot S}$, a quantity that becomes time-independent after an initial transient period.

One can define a dimensionless compressibility  $C = \frac{\langle(\vec\nabla_2 \cdot \vec v)^{2}\rangle}{\langle(\vec\nabla_2\vec v)^{2}\rangle}$ \cite{robNJP2004},  which lies between 0 and 1 for an isotropic system.  Previous experiments and numerical studies \cite{robNJP2004} have consistently reported $C = 0.5$ at the surface, thus making the system of floaters a strongly compressible one.

\section{Experiment}
The experimental system is a tank of water 1 m x 1 m in lateral dimensions, filled to a depth of 30 cm and maintained in a turbulent steady state. The experimental setup is discussed in detail in \cite{robNJP2004}. Turbulence is generated by an 8 hp pump that circulates water in the tank via an array of 36 rotating capped jets situated at the tank floor. The points of turbulent injection are therefore far removed from the surface where measurements are made. The injection scheme was chosen to minimize the amplitude of waves on the surface \cite{goldburg2001}. Hollow glass spheres of mean diameter 50 $\mu$m and specific gravity 0.25 follow the local surface flow. A beam from a diode-pumped laser (5.5 W) is passed through a cylindrical lens to generate a sheet of light that illuminates the surface. Light scattered by the particles is captured by a high speed camera (Phantom v5.0) that records  particle positions and velocities at 100 frames per second. Steady state flow measurements are achieved by constantly seeding floaters from the bottom of the tank to compensate for those that are lost from the camera's field of view. The computer-stored record is broken up into a series of images that are fed into a particle tracking program in consecutive pairs to obtain the experimental steady-state flow velocity fields. A total of 2040 instantaneous velocity fields spanning a duration of 20 s are obtained. There are on the average 25000 velocity vectors in each velocity field, providing reliable spatial resolution over a square area of side length L = 9.3 cm. Parameters that characterize turbulence on the surface are listed in table~\ref{table1}.

\section{Analysis}
 For the analysis discussed below, the experimentally obtained velocity fields are seeded with fictitious (surrogate) particles via computer programming. At $t = 0$, the 1024 x 1024 pixel grid of the velocity field is decorated with a uniform array of surrogate particles 6 pixels apart, providing $170^2 = 28900$ surrogate particles. Evolution of this initially uniform particle distribution is dictated by the experimentally obtained velocity fields. The particles are tracked by the program instant-by-instant in the lagrangian frame. Though some particles leave the field of view during the evolution time, they are tracked as long as they remain in the field of view. Essential to the theory of FF is the requirement that the number of particles in the system be conserved. Hence for every surrogate particle that leaves the field of view, a new particle is introduced at a random spatial point within it, thereby representing a new lagrangian trajectory.

The focus of the experiments is on the  rate of entropy production at times $t$ large compared to the turnover time of the largest eddies, estimated to be 0.54 s. The 20 s time record is broken up into 20 time-uncorrelated data sets of 1 s duration. The surrogate particles are tracked and the two terms of eq.~\ref{sdot1} are measured for each data set and averaged over the 20 sets. Figure~\ref{fourclouds} shows the time evolution of one of these sets.  Surrogate particles are used for entropy rate measurements since they allow an initial periodic placement which is not possible with real particles. In addition the surrogates also permit particle tracking in the lagrangian frame of reference. Lagrangian tracking of real particles is beyond the experimental capability of the current scheme.

\section{Discussion}
Before considering the two terms in eq.~\ref{sdot1}, it is of interest to examine the time dependence of the entropy itself. Each instantaneous snapshot is divided into two-dimensional bins (cells) of side length 8 pixels. The particle concentration $n_{i}(t)$ is the number of particles in the $i^{th}$ cell divided by the total number of particles in the field of view at time $t$.  The local entropy is calculated for each cell $i$ and the total instantaneous entropy as expressed in eq.~\ref{S} is obtained from an average over all cells $S(t) = -\sum_{i=1}^{N}n_{i}(t)~{\ln}~n_{i}(t)$ in a given snapshot. Here N is the total number of cells in the field of view.

\begin{figure}
\onefigure[width=2.5in]{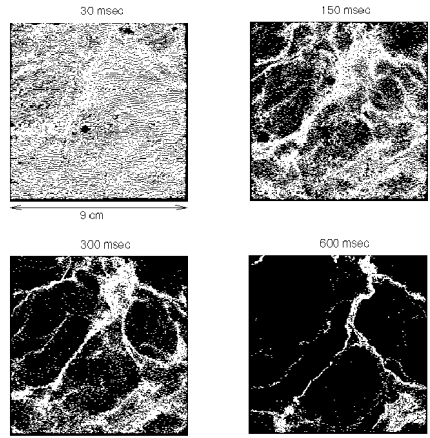}
\caption{A system of fictitious (surrogate) particles is introduced on the experimental velocity fields at $t = 0$. The distribution of particles is homogeneous to start with. However the particles quickly flee regions of fluid up-wellings and cluster into thin ribbon like structures around fluid down-wellings as time progresses. The evolution of clusters is shown for four snapshots in time as they evolve from a nearly homogeneous distribution at $t = 30$ ms towards an inhomogeneous distribution through $t = 150$ ms and 300 ms. The experiment is almost over at $t = 600$ ms by which time particles have almost completely clustered into ribbon like structures.}
\label{fourclouds}
\end{figure}

\begin{figure}
\onefigure[width=2.5in]{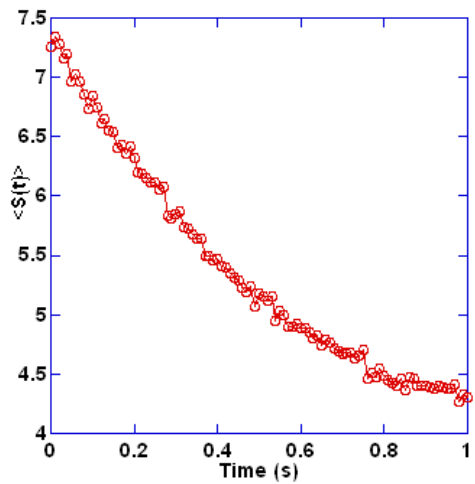}
\caption{Time trace of the entropy S(t). The angular brackets around $S(t)$ denote an ensemble average over the 20 time-uncorrelated sets. This quantity shows a monotonic decrease in time arising from the clustering of particles at the surface as observed in fig.~\ref{fourclouds}.}
\label{entropy}
\end{figure}

Figure~\ref{entropy} shows the time evolution of $S(t)$ over an interval of 1 s. The curve is an ensemble average of the 20 data sets measured at each instant of time $t$.  Observe that $S(t)$ from eq.~\ref{S} decreases monotonically through this time interval, implying an increase in particle clustering.  This accords  with  the visual observations of  clustering  tracked over 0.6 s in  fig.~\ref{fourclouds}. The ensemble of  $S(t)$ measurements is not large enough to completely average out its temporal fluctuations, making it  impossible to  extract its derivative.  That function is best obtained using eq.~\ref{sdot1}.

The velocity divergence field $\omega({\bf r},t)$ is obtained by taking the component-wise spatial derivative of the velocity fields at each instant. With the simultaneous measurement of local particle density $n({\bf r},t)$, one has all  the information needed to measure the integrands in eq.~\ref{sdot1}. This spatial and temporal record of velocity divergence and particle concentrations is coupled together to obtain the area term of eq.~\ref{sdot1}. The boundary term is obtained from particle concentrations at the periphery as they leave the field of view. This information about the concentration, when coupled with the velocity component perpendicular to the periphery, provides the boundary term. 

\begin{figure}
\onefigure[width=2.5in]{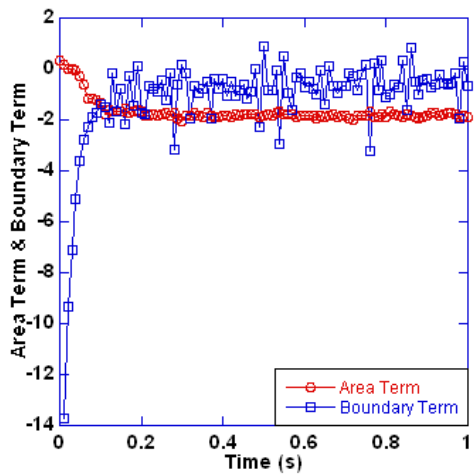}
\caption{The production rate of entropy ($\dot S$) in eq.~\ref{sdot1}. The area Term (red circles) and Boundary Term (blue squares) in eq.~\ref{sdot1} reach steady state (~0.2 s) within a fraction of the large eddy turnover time of turbulence (~0.54 s).}
\label{sdot}
\end{figure}

Figure~\ref{sdot} shows both the area term (open circles) and the boundary term (open squares) of $\dot S$  in units of s$^{-1}$. These two curves  are the primary results of this study. It is seen from fig.~\ref{sdot} that the dominant contribution comes from the area term, on which FF focus. After the system has reached a steady state (in roughly  0.2 s), the area and boundary terms reach values of -1.82 $\pm$ 0.07 Hz and -0.60 $\pm$ 0.7 Hz respectively. It is apparent that  the boundary term is hardly distinguishable from zero though its uncertainty is appreciable.

The data in fig. \ref{sdot} are robust to changes in the cell size ($m_c$) provided that change is kept within a reasonable range. The determination of velocity divergence requires calculation of spatial derivatives which cannot be reliably computed below $m_c = 8$ pixels. This sets the lower bound on $m_c$ for this experiment. The mean values of $\dot S$ evaluated at $m_c$ = 8 pixels and $m_c$ = 16 pixels were within a standard deviation. On the other hand for $m_c$ = 32 pixels, the mean $\dot S$ dropped measurably. Normally the velocity divergence should be computed at the smallest scales of turbulence, namely the Kolmogorov scale $\eta$. This scale cannot be resolved in this experiment due to limitations of camera resolution. A systematic error is expected since the cell size of $m_c$ = 8 pixels strictly falls in the inertial range of turbulence. It is therefore surprising that the measured values of $\dot S$ are robust for $m_c$ = 8 and 16 pixels despite spatial derivatives being computed at scales larger than the Kolmogorov scale $\eta$. Also analyzed was the dependence of $\dot S$ on change in area. No significant change in the area term was observed when the field of view was reduced by a factor of 2.

\begin{figure}
\onefigure[width=2.5in]{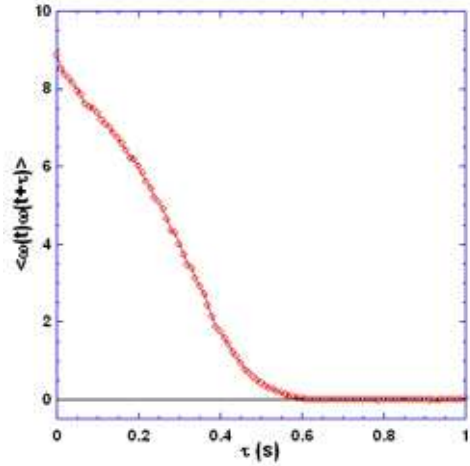}
\caption{Temporal correlation $C_{div}(\tau)$ of the lagrangian velocity divergence  $\omega({\bf r},t) = \vec\nabla_2 \cdot {\bf v}({\bf r},t)$ vs. time $\tau$. It was calculated individually for velocity divergence time traces of 28900 lagrangian trajectories that formed the initial uniform distribution, and averaged over to obtain the plot in the figure.}
\label{divcorr}
\end{figure}

The steady-state value of the area term cannot be predicted from general considerations. However, if the system is highly chaotic, in the sense that it obeys SRB statistics ~\cite{eckmann1985,dorfman1999,grishaNJP2004}, then the area term in Eq. \ref{sdot1} can be ascertained from Eq. \ref{sdot2}. The measured correlation function of lagrangian velocity divergence (Eq. \ref{sdot2}) is shown in fig.~\ref{divcorr}. It was extracted from an average over the 28900 lagrangian trajectories that formed the initial uniform distribution of surrogate particles at time $t = 0$ s. This uniform distribution ensures an equal weighting over all sources (fluid up-wellings) and sinks (fluid down-wellings) of the flow field. The area under this correlation function with a negative prefactor was calculated to be -2.4 $\pm$ 0.02 Hz. It is likelythat the source of the slightly higher value obtained from the correlation function comes from the boundary contribution which yields  -0.6 $\pm 0.7$ Hz.  It cannot be ruled out, however, that the discrepancy is controlled by the experimental impossibility of making measurements down to the dissipative scale $\eta$.  Whatever the reason, the two values of this entropy rate differ from the average value by less than 15\%.

The two Lyapunov exponents have not been measured in this experiment, but the ribbon-like clusters in fig.~\ref{fourclouds} suggest one exponent is negative and the other positive. For SRB statistics, one expects smooth behavior of the particle density along the unstable direction (where $\lambda_1 > 0$) and a fractal distribution of $n({\bf r})$ along the transverse direction corresponding to $\lambda_2 < 0$. These two exponents have been extracted from a computer simulation of clustering on the surface of a turbulent fluid, the starting point being the Navier-Stokes equation ~\cite{boffetta2004}. In that study, the parameters were close to those of the present experiments. This study yields $\lambda_1$ = +0.3 Hz  and $\lambda_2$ = -2.0 Hz. The exponents have opposite sign, as expected. Their sum, -1.7 Hz, agrees well with the measured value for the area term, $\dot S_A$ = -1.82 $\pm$ 0.07 Hz.

\section{Summary}
In summary, an experiment has been described in which the the rate of entropy change has been measured for a compressible system of particles that float on a turbulent fluid.  In this steady state experiment it is observed that ${\dot S}$ reaches a time-independent value in a fraction of the lifetime of the largest eddies in the underlying flow. Also measured is the functional form of the lagrangian velocity divergence correlation function, for which no theoretical prediction exists. Assuming the density distribution of the floaters admits SRB statistics, ${\dot S}$ should be equal to the sum of two Lyapunov exponents, which have been measured in a computer simulation carried under conditions very similar to those of the present experiment. It is hoped that the statistical approach to turbulence taken by Falkovich and Fouxon and others ~\cite{goldenfeld2006,grisha2006} will complement the more traditional approach, where the starting point is the Navier-Stokes equation.

\begin{table}
\caption{Parameters of compressible turbulence measured at the surface.}
\label{table1}
\begin{center}
\begin{tabular}{|l|l|l|}
\hline
Parameter  & Expression  & Measured value \\ \hline
Taylor microscale $\lambda$ (cm) & $\sqrt{\frac{v_{rms}^{2}}{\langle ({\partial v_{x}}/{\partial x})^{2}\rangle}}$ & 0.3\\
$Re_{\lambda}$  & $\frac{v_{rms}\lambda}{\nu}$  & 93\\
Integral Scale $l_{0}$ (cm) & $\int dr \frac{\langle v_{||}(x+r)v_{||}(x)\rangle}{\langle (v_{||}(x))^{2}\rangle}$ & 1.2\\
Dissipation Rate $\varepsilon_{diss}$ $(cm^{2}/s^{3})$  & $10\nu\langle (\frac{\partial v_{x}}{\partial x})^{2}\rangle$  & 10.7\\
Kolmogorov Scale $\eta$ (cm) & $\eta = (\frac{\nu^3}{\varepsilon})^{1/4}$ & 0.02\\
Large Eddy Turnover Time $\tau_{l_{0}}$ (s) & $\tau_{l_{0}} = \frac{l_{0}}{v_{rms}}$ & 0.54\\
RMS Velocity $v_{rms}$ (cm/s) & $v_{rms} = \sqrt{\langle v_{||}^{2}\rangle - \langle v_{||} \rangle^{2}}$ & 2.6\\ \hline
\end{tabular}
\end{center}
\end{table}

\acknowledgments
MMB and WIG acknowledge helpful discussions with G. Falkovich, A. Pumir, N. Goldenfeld, G. Gallavotti, K. Gawedzki and X.-L. Wu. This work was supported by the National Science Foundation under grant number DMR-0201805.

\end{document}